\documentclass[aps,nofootinbib,prd,superscriptaddress,tightenlines,notitlepage,twocolumn,showpacs,floatfix]{revtex4-1}
\usepackage[colorlinks]{hyperref}
\usepackage{tabularx}
\usepackage{bm}
\usepackage{graphicx}
\usepackage{amssymb,bm,tensor}
\usepackage{xcolor}
\usepackage{amsmath}
\usepackage[varg]{txfonts}
\usepackage{enumerate}
\usepackage[normalem]{ulem}
\usepackage{bm}% bold math
\usepackage[OT1]{fontenc}
\usepackage{threeparttable}
\usepackage{mathtools}
\usepackage{soul,xcolor}

\usepackage{scalerel,stackengine}
\stackMath
\newcommand\reallywidehat[1]{%
\savestack{\tmpbox}{\stretchto{%
  \scaleto{%
    \scalerel*[\widthof{\ensuremath{#1}}]{\kern-.6pt\bigwedge\kern-.6pt}%
    {\rule[-\textheight/2]{1ex}{\textheight}}%WIDTH-LIMITED BIG WEDGE
  }{\textheight}%
}{0.5ex}}%
\stackon[1pt]{#1}{\tmpbox}%
}

\begin{document}

\setstcolor{red}

\title{Ringdown mode amplitudes of charged binary black holes}

%\date{\today}

\author{Zexin Hu}%\email{huzexin@pku.edu.cn}
    \email{huzexin@pku.edu.cn}
\affiliation{Department of Astronomy, School of Physics, Peking University, Beijing 100871, China}
\affiliation{Kavli Institute for Astronomy and Astrophysics, Peking University, Beijing 100871, China}
\affiliation{Theoretical Astrophysics, Eberhard Karls University of T\"ubingen, 72076 T\"ubingen, Germany}

\author{Daniela D. Doneva}
\affiliation{Departamento de Astronom\'ia y Astrof\'isica, Universitat de Val\`encia,
Dr. Moliner 50, 46100, Burjassot (Val\`encia), Spain}
\affiliation{Theoretical Astrophysics, Eberhard Karls University of T\"ubingen, 72076 T\"ubingen, Germany}

\author{Ziming Wang}
\affiliation{Department of Astronomy, School of Physics, Peking University, Beijing 100871, China}
\affiliation{Kavli Institute for Astronomy and Astrophysics, Peking University, Beijing 100871, China}

\author{Vasileios Paschalidis}
\affiliation{Department of Astronomy, University of Arizona, Tucson, AZ, USA}
\affiliation{Department of Physics, University of Arizona, Tucson, AZ, USA}

\author{Gabriele Bozzola}
\affiliation{AWS Center for Quantum Computing, Pasadena, 91125}
\email{Work done prior to joining AWS.}

\author{Stoytcho S. Yazadjiev}
\affiliation{Department of Theoretical Physics, Sofia University ``St. Kliment Ohridski'', 5 J. Bourchier Blvd. Sofia 1164, Bulgaria}
\affiliation{Institute of Mathematics and Informatics, Bulgarian Academy of Sciences, Acad. G. Bonchev St. 8, Sofia 1113, Bulgaria}

\author{Lijing Shao}
\affiliation{Kavli Institute for Astronomy and Astrophysics, Peking University, Beijing 100871, China}
\affiliation{National Astronomical Observatories, Chinese Academy of Sciences, Beijing 100012, China}

\begin{abstract}

The ringdown phase of the binary black hole (BBH) merger provides a clean and direct probe of strong-field gravity and tests of the nature of black holes. The quasinormal mode (QNM) frequencies in modified gravity theories, as well as their amplitudes and phases, might deviate from the Kerr ones in general relativity. Charged black holes (BHs) in Einstein-Maxwell theory provide an excellent example of a beyond-Kerr solution with direct astrophysical and fundamental physics applications. In this work, we extract the ringdown mode amplitudes and phases for charged BBH mergers based on fully general relativistic simulations with charge-to-mass ratio up to $0.3$. Our results suggest that even though the inspiral phase of charged BBHs can be significantly accelerated or decelerated, the ringdown mode excitation only changes mildly. We further explore the charge detectability with the ringdown-only signal for the Einstein Telescope and Cosmic Explorer. We find that previous studies may have overestimated the charge detectability and including higher modes in charged waveforms is necessary for future ringdown analysis. This constitutes the first such analysis based on waveforms generated by numerical relativity simulations of charged BHs in full Einstein-Maxwell theory. 

\end{abstract}

\maketitle

%% === main body of the paper ===

\allowdisplaybreaks

%---------------------------------------------------------------------
\section{Introduction}
%---------------------------------------------------------------------

The detection of the first gravitational wave (GW) event GW150914 opened a new area of GW astronomy~\cite{LIGOScientific:2016aoc}. GWs generated by the coalescence of compact binaries provide information for studying strong-field gravitation~\cite{LIGOScientific:2016lio,LIGOScientific:2021sio}, compact stars~\cite{LIGOScientific:2018cki,KAGRA:2021duu}, and cosmology~\cite{LIGOScientific:2017adf}. The final stage of a binary black hole (BH) merger, known as ringdown, is a unique probe of the nature of BHs~\cite{Isi:2019aib,Isi:2020tac,Cardoso:2016rao}. In the ringdown stage, before the late-time tail~\cite{Price:1971fb}, the GW signal can be described by a sum of quasinormal modes (QNMs), which are damped sinusoids~\cite{Teukolsky:1972my,Teukolsky:1973ha}. As a consequence of the no-hair theorem in general relativity  (GR)~\cite{Israel:1967wq,Carter:1971zc,Robinson:1975bv}, the complex frequencies of these QNMs depend only on the mass and spin for an astrophysical uncharged BH. Measuring the frequencies of QNMs thus provides a clean and direct test of gravity theories via BH spectroscopy~\cite{Dreyer:2003bv,Berti:2025hly}. In recent years, there has also been a growing interest in the amplitudes and phases of the QNMs~\cite{Yang:2017zxs,Cheung:2023vki,Mitman:2025hgy,Nobili:2025ydt}, which contain information about the initial perturbations excited during the plunge and merger phases~\cite{Berti:2006wq}. As a theory should consistently provide predictions for all stages of a GW event, one can also perform a gravity test by comparing the relative excitation of QNMs with theoretical predictions~\cite{Berti:2007fi,Kamaretsos:2011um,Kamaretsos:2012bs}. However, the study of QNMs excitation requires extracting QNM amplitudes and phases from numerical simulations, which is challenging due to the ill-defined nature of the problem~\cite{Cheung:2023vki,Mitman:2025hgy,Giesler:2024hcr}: a small amount of noise in the data can lead to much larger errors when fitting a sum of exponentials with unknown frequencies and unknown numbers. Also, studying QNM excitations in theories beyond GR can currently be performed reliably only in a very limited range of theories that have a well-posed formalism~\cite{East:2020hgw,Kovacs:2020ywu,Kovacs:2020pns,AresteSalo:2022hua,Rubio:2023eva}.

Charged BHs in Einstein-Maxwell theory provide an excellent springboard for studying the ringdown features of beyond-Kerr BHs, while also having direct astrophysical and fundamental physics applications, such as modeling certain types of modified gravity, minicharged dark matter, and dark electromagnetism, among others~\cite{Bozzola:2020mjx}. The most generic BH solution that is stationary and asymptotically flat in this theory is the well-known 
Kerr-Newman (KN) solution, which is uniquely characterized by the mass, spin, and electric charge of 
the BH~\cite{Newman:1965my,Mazur:1982db}. Although astrophysical BHs are expected to
carry only negligible electric charge due to various neutralization mechanisms~\cite{Gibbons:1975kk,Eardley:1975kp,Blandford:1977ds,Gong:2019aqa}, there is little
observational support for this expectation~\cite{Zajacek:2018ycb,Zajacek:2018vsj}.
Therefore, whether all astrophysical BHs have negligible charge or not remains a question to be answered by observations. Furthermore, proposed
discharge mechanisms almost exclusively apply to the electric charge, while theories beyond Standard Model or GR permit BHs to carry different kinds of charges. For example, BHs could carry magnetic charge by 
primordial magnetic monopoles~\cite{Preskill:1984gd}; minicharged dark matter model, which predicts particles with a lower charge-to-mass ratio, could evade some discharge mechanisms~\cite{Cardoso:2016olt}; modified gravity 
theories, including additional vector fields like the bumblebee gravity theory, also allow charged BHs~\cite{Kostelecky:2003fs,Xu:2022frb}. See also Ref.~\cite{Bozzola:2020mjx} for additional discussions. At the physical scale of BH mergers, many of the charges mentioned above  have similar effects and can be parametrized with the same parameter as in the Einstein-Maxwell theory. Thus, studying charged BHs can provide insight into these theories. 

The ringdown-only analysis of current GW observations only gives weak constraints on the charge-to-mass ratio due to the limited signal-to-noise ratios (SNRs) for postmerger data~\cite{Carullo:2021oxn,Gu:2023eaa}. Next-generation GW detectors, such as the ground-based Einstein Telescope (ET)~\cite{Punturo:2010zz,Hild:2010id,Abac:2025saz} and Cosmic Explorer (CE)~\cite{Reitze:2019iox,Reitze:2019dyk}, or the space-base detectors Laser Interferometer Space Antenna (LISA)~\cite{LISA:2017pwj}, Taiji~\cite{Hu:2017mde}, and TianQin~\cite{TianQin:2015yph} are expected to detect GWs in a different frequency band with much higher sensitivity. Existing work suggests that one could constrain the charge-to-mass ratio to less than $0.2$ for a GW150914-like event with future detectors~\cite{Gu:2023eaa}.

In this work, we focus mainly on two aspects. Based on existing simulations of charged binary BH (BBH) mergers~\cite{Bozzola:2021elc}, we perform QNM extraction to analyze the dependence of ringdown amplitudes and phases on the charge-to-mass ratio of the progenitor BHs. Our results suggest that QNM excitations only mildly depend on the charge-to-mass ratio up to a value of 0.3, which can already have a strong deviation from GR in the inspiral phase. We also perform the first Bayesian analysis of charge detectability using waveforms generated by simulations of charged BBHs in full Einstein-Maxwell theory. Assuming the ET's or CE's designed power spectral density~\cite{Yi:2021wqf} and GW150914-like event~\cite{LIGOScientific:2016aoc,LIGOScientific:2016vlm}, the subsequent analysis indicates that previous studies may overestimate the charge detectability and including higher modes in charged waveforms is necessary for future ringdown analysis.

The remainder of this paper is organized as follows. In Sec.~\ref{sec:sim}, we briefly describe the numerical simulations that our study relies on, and in Sec.~\ref{sec:ext} we introduce our procedures for QNM extraction. The extraction results are presented in Sec.~\ref{sec:extraction result}. We discuss the charge detectability for future observations in Sec.~\ref{sec:det} and conclude our work in Sec.~\ref{sec:con}. Unless otherwise specified throughout we adopt geometrized units in which $G=c=1$.

%---------------------------------------------------------------------
\section{Simulations}\label{sec:sim}
%---------------------------------------------------------------------

Our work is based on the simulations presented in Refs.~\cite{Bozzola:2020mjx,Bozzola:2021elc}, which are quasi-circular inspiral and merger of charged, non-spinning BHs in full GR. The mass ratio of the system is chosen to be GW150914-like ($q=36/29$)~\cite{LIGOScientific:2016aoc,LIGOScientific:2016vlm}. The BHs are initially characterized by gravitational (quasilocal) mass $M_i$ and charge $Q_i$, with a charge-to-mass ratio $\lambda=Q_i/M_i$ varying between $\lambda=0$ and $|\lambda|=0.3$. There are three branches of simulations: the two initial BHs have the same charge-to-mass ratio and the charges have the same sign; only one of the initial BHs is charged; and the two initial BHs have the same charge-to-mass ratio but with opposite charge. In the remainder of the paper, we use $++$, $+0$, and $+-$ to distinguish the three branches of simulations. 

\begin{figure}[h]
  \centering
  \includegraphics[width=0.48\textwidth]{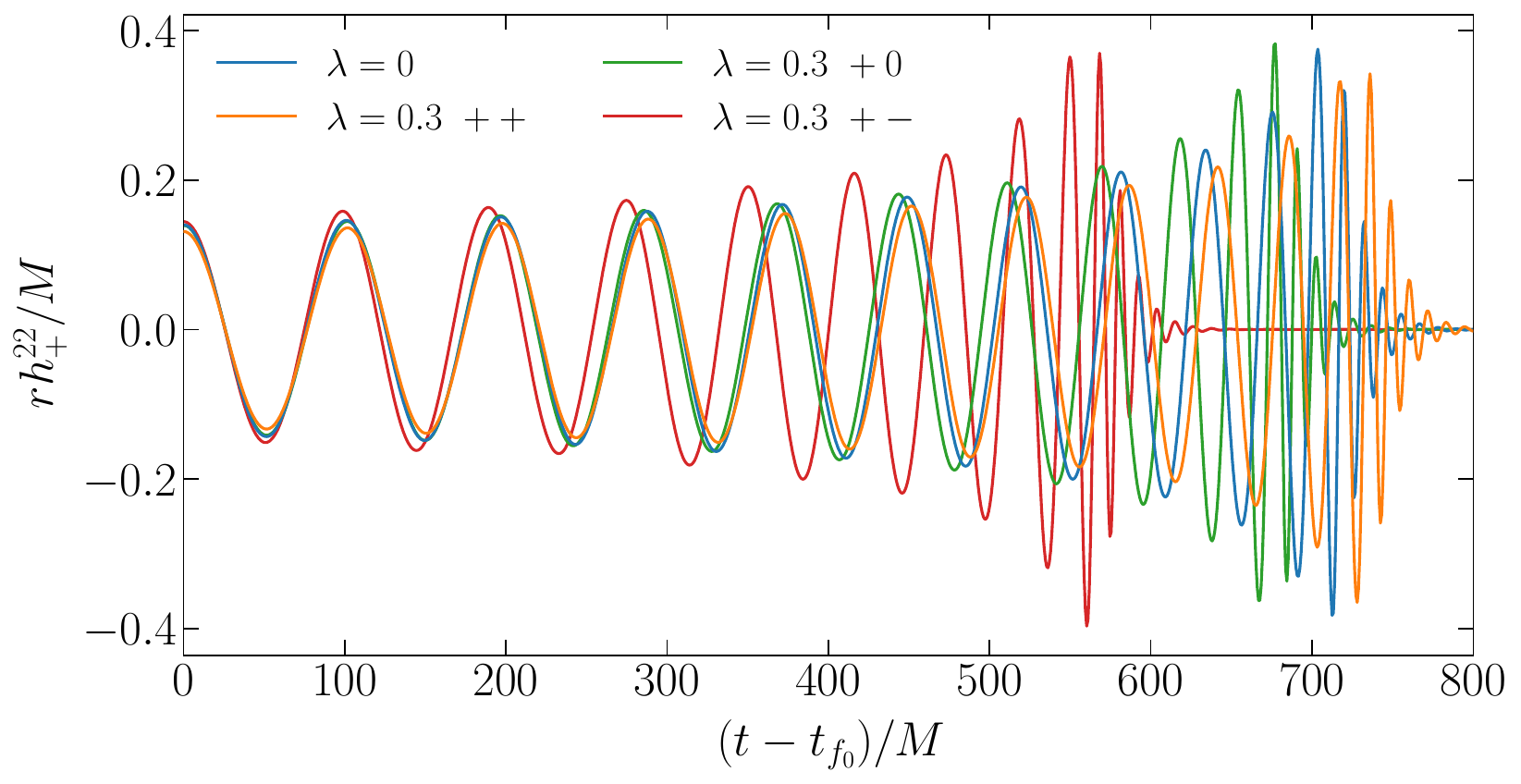}
  \caption{\label{fig:waveform} The $22$ mode of the GW plus polarization $h_+^{22}$ emitted by charged BBHs in three cases with $|\lambda|=0.3$. We also show a non-charged case for comparison. The waveforms are aligned at a specific GW frequency $Mf_0=9.61\times10^{-3}$, which is motivated by the LIGO sensitivity for detecting a GW150914-like event~\cite{Bozzola:2021elc}.}
\end{figure}

The simulations were performed with the {\tt EinsteinToolkit}~\cite{einsteintoolkit} infrastructure; the spacetime evolution was performed with the \texttt{Lean} code~\cite{Sperhake:2006cy}, which adopts the Baumgarte-Shapiro-Shibata-Nakamura (BSSN) formulation of Einstein's equations~\cite{Shibata:1995we,Baumgarte:1998te}, and the electromagnetic fields are evolved with the \texttt{ProcaEvolve} code~\cite{Zilhao:2015tya}. The initial data of charged BBH are generated by \texttt{TwoChargedPunctures} developed in Ref.~\cite{Bozzola:2019aaw}. Further details on how low eccentricity is achieved and the numerical setup can be found in Refs.~\cite{Bozzola:2019aaw,Bozzola:2020mjx,Bozzola:2021elc}. {\tt Kuibit}~\cite{kuibit} was used in the analysis of the simulations.
In Fig.~\ref{fig:waveform}, we show the 22 mode of GW plus polarization $h_+^{22}$ for three charged BBH simulations with $|\lambda|=0.3$ and a non-charged case for comparison. The waveforms are aligned at the frequency $Mf_0=9.61\times10^{-3}$~\cite{Bozzola:2021elc}. The plot demonstrates that during the inspiral, there can be substantial dephasing of the charged cases compared to the uncharged one, which can be used to place constraints on the BH charge~\cite{Bozzola:2019aaw}.

%---------------------------------------------------------------------
\section{Mode extraction}\label{sec:ext}
%---------------------------------------------------------------------

We follow the method proposed in Ref.~\cite{Cheung:2023vki} for extracting the QNM amplitude and phase information from the simulations but with some differences. We summarize our procedure in this section. 

In the simulation data we adopt, the GWs are extracted using a spin-weighted spherical harmonic decomposition of the Newman-Penrose scalar at a finite radius~\cite{Boyle:2019kee}
\begin{equation}
    \Psi_4=\sum_{l,m}\Psi_4^{lm}\prescript{}{-2}{Y}_{lm}\,,
\end{equation}
where $\prescript{}{-2}{Y}_{lm}$ are the spin-weighted $s=-2$ spherical harmonics. Thus, instead of fitting the GW strain $h=h_{+}-ih_{\times}$, we choose to fit $\Psi_4=\ddot{h}$ as it is the direct output of the simulations. The commonly adopted method for converting $\Psi_4$ to $h$, such as the fixed frequency integration method~\cite{Reisswig:2010di}, can have large uncertainties in the late ringdown phase that could affect our fitting procedure. We use the $\Psi_4$ extracted from a finite radius of $110.69\,M$, where $M$ is the total ADM mass of the initial system. We have examined that extrapolating the waveform to infinity using the standard method shown in, for example, Ref.~\cite{Boyle:2019kee} only gives a slightly smaller overall amplitude. The difference between the amplitude of the waveform at $110.69$M and the extrapolated waveform is smaller than the estimated extraction precision, as shown later. Considering the unknown errors introduced by the extrapolation procedure and the limited accuracy of the simulation waveforms, we  use the waveform extracted at $110.69$M. Future work applying the Cauchy characteristic extraction method can avoid  extrapolation errors, and provide improved waveforms~\cite{Bishop:1996gt,Taylor:2013zia,Giannakopoulos:2023zzm,Ma:2024hzq}.

We fit the spin-weighted spherical harmonic decomposition coefficients $\Psi_4^{lm}$ with the following ansatz
\begin{equation}\label{eq:fitting ansatz}
    \Psi_4^{lm} (t)=\sum_{k=1}^N B_k e^{-i(\omega_{r,k}+i\omega_{i,k})(t-t_{\rm peak})-i\varphi_k}\,,
\end{equation}
where  the amplitudes $B_k$ and the phases $\varphi_k$ are free parameters of our fits. The time $t_{\rm peak}$ is the reference time of the amplitude and the phase, and we set it to be the peak time of the strain $|h_{22}|$ to compare with the results in Ref.~\cite{Cheung:2023vki}. Although, as mentioned before, the conversion from $\Psi_4$ to $h$ can be inaccurate at the very beginning and the late time of the waveform, the bulk part of the waveform, including the peak, is less affected by the choice of parameters in the integration procedure and thus provides a reliable reference time~\cite{Reisswig:2010di}. Further, $B_k$ and $\varphi_k$ defined above are related to the amplitude $A_k$ and the phase $\phi_k$ of the GW strain~\cite{Cheung:2023vki}
\begin{equation}
    h_{lm}(t)=\sum_{k=1}^{N}A_k e^{-i(\omega_{r,k}+i\omega_{i,k})(t-t_{\rm peak})-i\phi_k}\,,
\end{equation}
for related QNMs via 
\begin{equation}
    B_k e^{-i\varphi_k}=-(\omega_{r,k}+i\omega_{i,k})^2A_k e^{-i\phi_k}\,.
\end{equation}

Equation~\eqref{eq:fitting ansatz} is a superposition of $N$ modes, as the coefficient $\Psi_4^{lm}$ will naturally contain the $lm0$ fundamental mode and all the $lmn$ overtones with $n\geq 1$~\cite{Baibhav:2023clw}. Moreover, there are other modes or not mode components from spherical-spheroidal mixing~\cite{Berti:2014fga}; quadratic modes~\cite{London:2014cma}; retrograde modes~\cite{Dhani:2020nik}; imperfect choice of the Bondi-van der Burg-Metzner-Sachs frame~\cite{Mitman:2022kwt}; non-linear response at the early time and numerical noise.
Nevertheless, one would expect that for $\Psi_4^{lm}$, the dominant mode should be the fundamental mode $lm0$ and for any $lmn$ mode, they should be mainly contained in $\Psi_4^{lm}$. Thus, for extracting a desired mode $lmn$, we only use the fitting result of $\Psi_4^{lm}$ with the same $l$ and $m$.

The QNM frequencies for KN BHs, calculated through a numerical solution of the perturbed field equations, are available in the literature but only for a limited number of QNM families~\cite{Dias:2021yju,Dias:2022oqm}. Specifically, in Ref.~\cite{Carullo:2021oxn}, fitting formulas for the $220,221,330$ modes are provided. Based on these modes, we are able to perform fittings to $\Psi_{4}^{22}$ and $\Psi_4^{33}$ that explicitly contain desired modes with fixed frequencies. However, for other modes with unknown spectrum, we can only do frequency-agnostic fittings where $\omega_{r,k}$ and $\omega_{i,k}$ in Eq.~\eqref{eq:fitting ansatz} are also regarded as free parameters. 

We have to note that even for $lm$ combinations with known fundamental mode frequencies \cite{Dias:2021yju,Dias:2022oqm}, the fitting is more complicated.  As we discussed above, each $\Psi_4^{lm}$ contains the dominant mode, which is usually $lm0$, as well as higher overtones and other components. Thus, for the fitting, we always include $N$ components. Except one or two of them having fixed frequency known from perturbative studies \cite{Dias:2021yju,Dias:2022oqm}, other components all have frequencies as free parameters and our fittings in total may have $4N$, $4N-2$ (for $\Psi_4^{33}$ as the frequency of the $330$ mode is known), or $4N-4$ (for $\Psi_4^{22}$ as the frequencies of the $220$ and $221$ modes are known) free parameters. 

The number of total modes contained in the fitting, $N$, does have impact on the final extraction results of the desired modes, even when the desired modes have fixed frequencies~\cite{Cheung:2023vki}. In general, including more modes in the fitting will lead to a more ``stable'' result according to the criterion introduced later, especially for the fitting including the early ringdown part. However, including too many modes might overfit the simulation waveform and requires more computational resources. Also, the very high-dimensional global minimization might be problematic. According to our experiments, we always start with $N=5$ and sometimes slightly adjust $N$ to find a more stable fitting, as discussed later.
 
The presence of a specific QNM is confirmed by the criterion that the mode amplitude and phase should be stable irrespectively of the time window we used in the fitting at least for some intervals~\cite{Cheung:2023vki}. Thus, for each component $\Psi_4^{lm}$, we perform a series of fittings using waveform data from $t_0$ to $t_{\rm end}$. We fix $t_{\rm end}=t_{\rm peak}+100\,M$ and vary $t_0$ from $t_{\rm peak}$ to $t_{\rm peak}+50\,M$. For each starting time $t_0$, the data are least-square fitted by Eq.~(\ref{eq:fitting ansatz}) with the trust-region reflective method provided in the \texttt{JaxFit} package~\cite{hofer:2022}. The final values of the QNM parameters are then defined to be the median value in the ``most stable'' time window. Following Ref.~\cite{Mitman:2025hgy}, we define the instability of a mode with fixed frequency in a given time window to be 
\begin{equation}
    \sigma_{f}=\sqrt{\left(\frac{\Delta {\rm Re}[C_{\rm QNM}]}{|C_{\rm QNM}|}\right)^2+\left(\frac{\Delta {\rm Im}[C_{\rm QNM}]}{|C_{\rm QNM}|}\right)^2}\,,
\end{equation}
where $C_{\rm QNM}=B_ke^{-i\varphi_k}$ is the complex amplitude of the mode, and $\Delta$ represents the standard deviation. For a mode that is fitted in a frequency-agnostic way, one also needs to ensure that the mode frequency is a constant. Therefore, we further consider the variation in the mode frequency as
\begin{equation}
    \sigma_{d}=\sqrt{\sigma_f^2+\frac{(\Delta \omega_r)^2+(\Delta \omega_i)^2}{\omega_r^2+\omega_i^2}}\,.
\end{equation}
The instabilities $\sigma_{f/d}$ defined above clearly depend on the length of the time window we used to evaluate them. Thus, they do not truly represent the uncertainties of the fitting results. Considering that the QNMs are damped exponentially, we choose the length of the time window for a desired $lmn$ mode to depend on its mode damping time~\cite{Mitman:2025hgy}. More specifically, we set the length of the time window for $lmn$ mode to be 
\begin{equation}\label{eq:time_window}
    \tau_{lmn}=-\frac{\ln 10}{ \omega_{i,\,{\rm Kerr},\,lmn}}\,,
\end{equation}
where $\omega_{i,\,{\rm Kerr},\,lmn}$ is the imaginary part of the QNM frequency of a Kerr BH that has the same spin as the KN BH we considered but with no charge. 
For agnostic fitting, we also require the final fitted mode frequency to be at least inside a given range around its Kerr counterpart as we expect that the deviation from the Kerr frequencies to be small even for $\lambda=0.3$. More specifically, we require
\begin{equation}
    \left|\omega-\omega_{\rm Kerr}\right|<\epsilon_\omega|\omega_{\rm Kerr}|\,,
\end{equation}
where $\omega$ is the fitted frequency and $\omega_{\rm Kerr}$ is the Kerr counterpart. We take $\epsilon_\omega=0.2$. 

%---------------------------------------------------------------------
\section{Extraction results}\label{sec:extraction result}
%---------------------------------------------------------------------
Based on the extraction procedure introduced in the previous section, here we present the results for our QNM extraction for charged BBH merger waveforms. 

We mainly focus on the amplitude and phase extraction of the $220$, $221$, and $330$ modes for which we can do a fixed frequency fitting, based on the results in Refs.~\cite{Dias:2021yju,Dias:2022oqm}. For the $220$ and $221$ modes, we fit $\Psi_4^{22}$ with two modes having fixed frequency simultaneously. We set a stability threshold of $\sigma_{\rm max}=0.2$ as in~\cite{Cheung:2023vki}. If the time window with minimum $\sigma_f$ still has $\sigma_f>\sigma_{\rm max}$, we adjust the total mode number $N$ to find a more stable fitting that satisfies the threshold, otherwise, we use $N=5$.

\begin{figure}[h]
  \centering
  \includegraphics[width=0.45\textwidth]{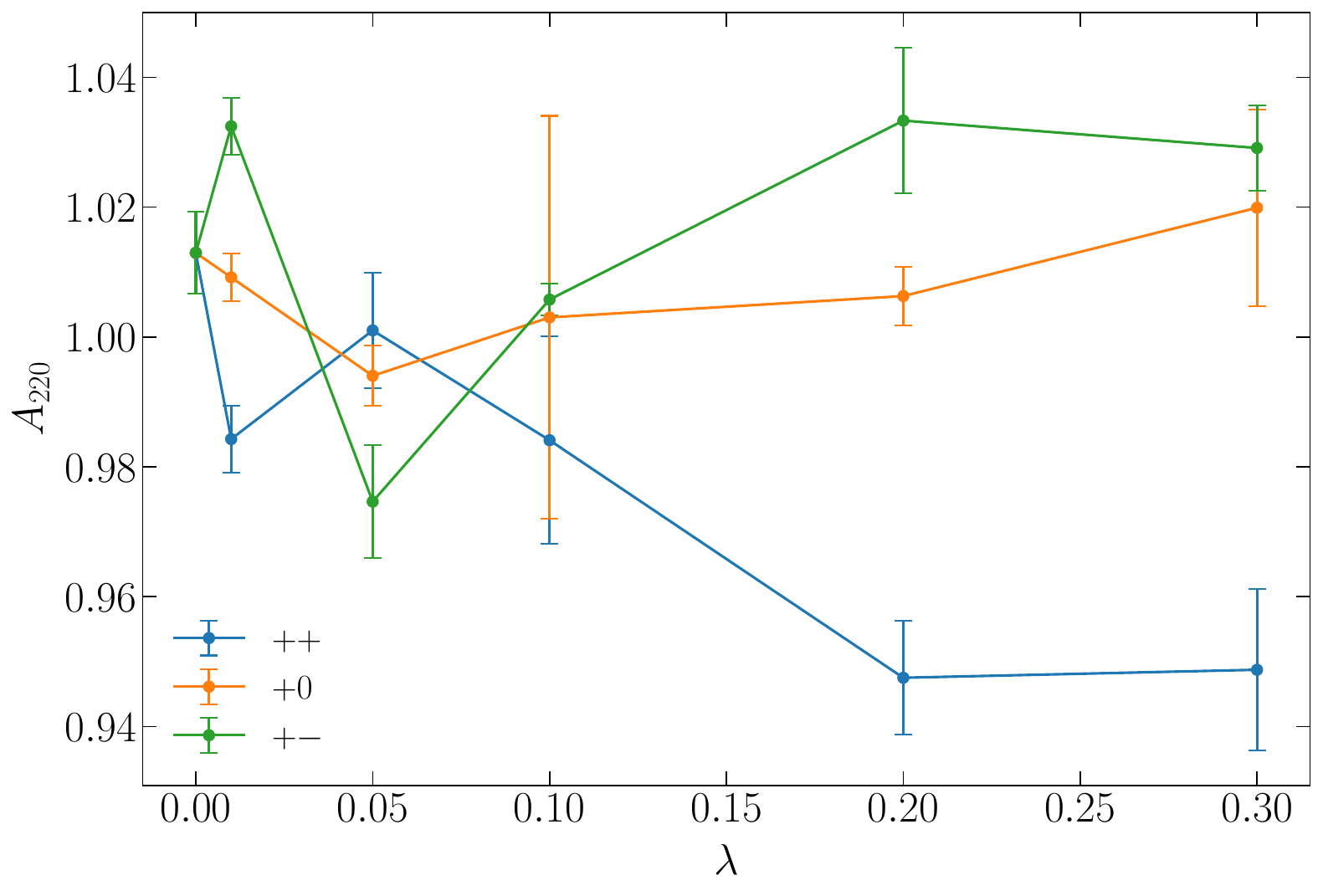}
  \caption{\label{fig:220_A} The extracted $220$ mode amplitude as a function of the charge-to-mass ratio $\lambda$, for the three families of 
  simulations, $++$, $+0$ and $+-$, discussed in the main text. The estimated uncertainty of the extracted amplitude is represented by the error bar. Note that $\lambda$ is the charge-to-mass ratio of the progenitor BHs. For the remnant BHs, the largest charge-to-mass ratio $\lambda_f$ for $++$, $+0$ and $+-$ simulations are around $0.3$, $0.15$ and $0$ respectively.}
  
\end{figure}

\begin{figure}[h]
  \centering
  \includegraphics[width=0.45\textwidth]{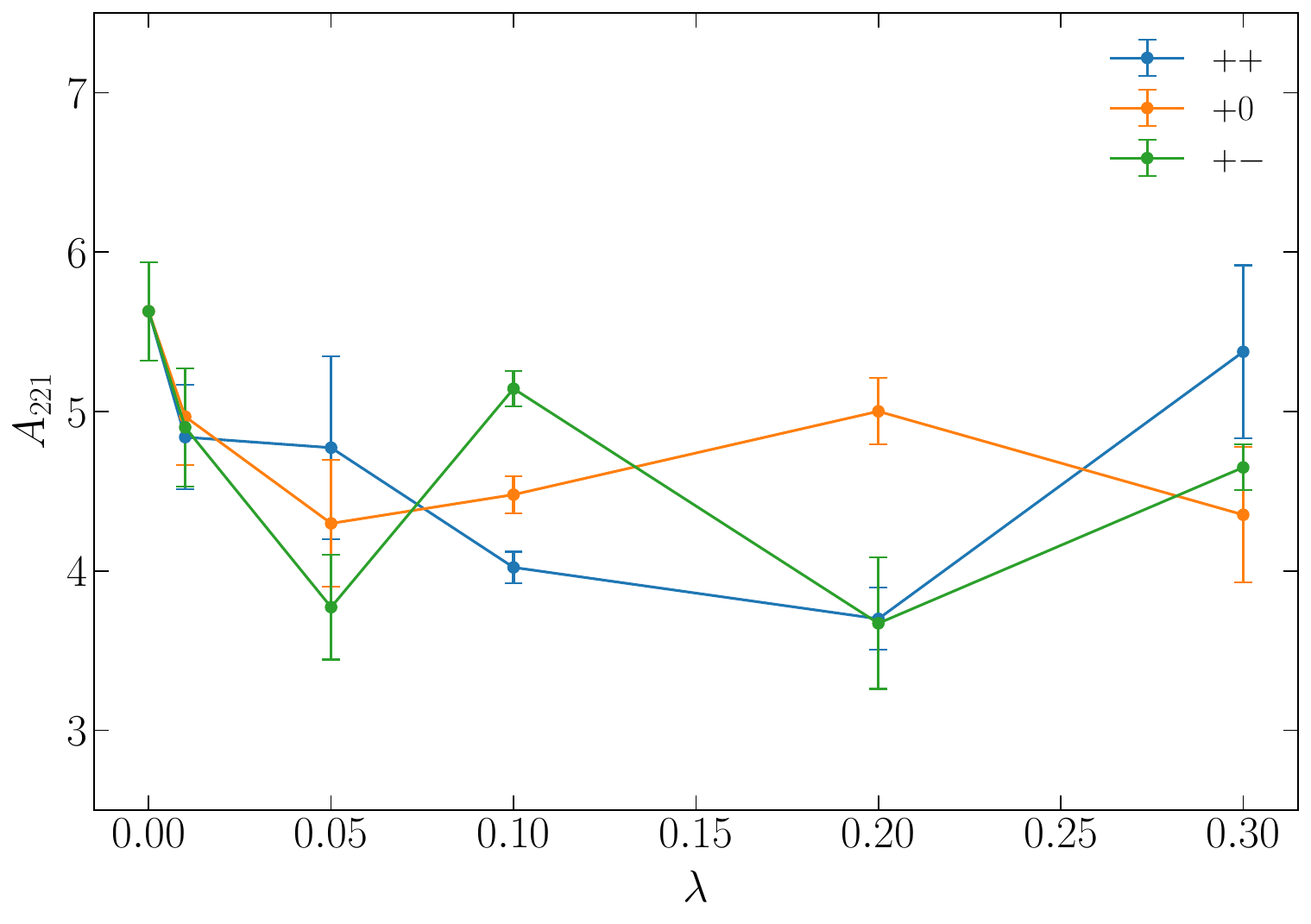}
  \caption{\label{fig:221_A} Similar to Fig.~\ref{fig:220_A}, but for $221$ mode. The estimated statistical uncertainty of the extracted amplitude may underestimate the real uncertainty as discussed in the main text.
  }
\end{figure}

\begin{figure}[h]
  \centering
  \includegraphics[width=0.45\textwidth]{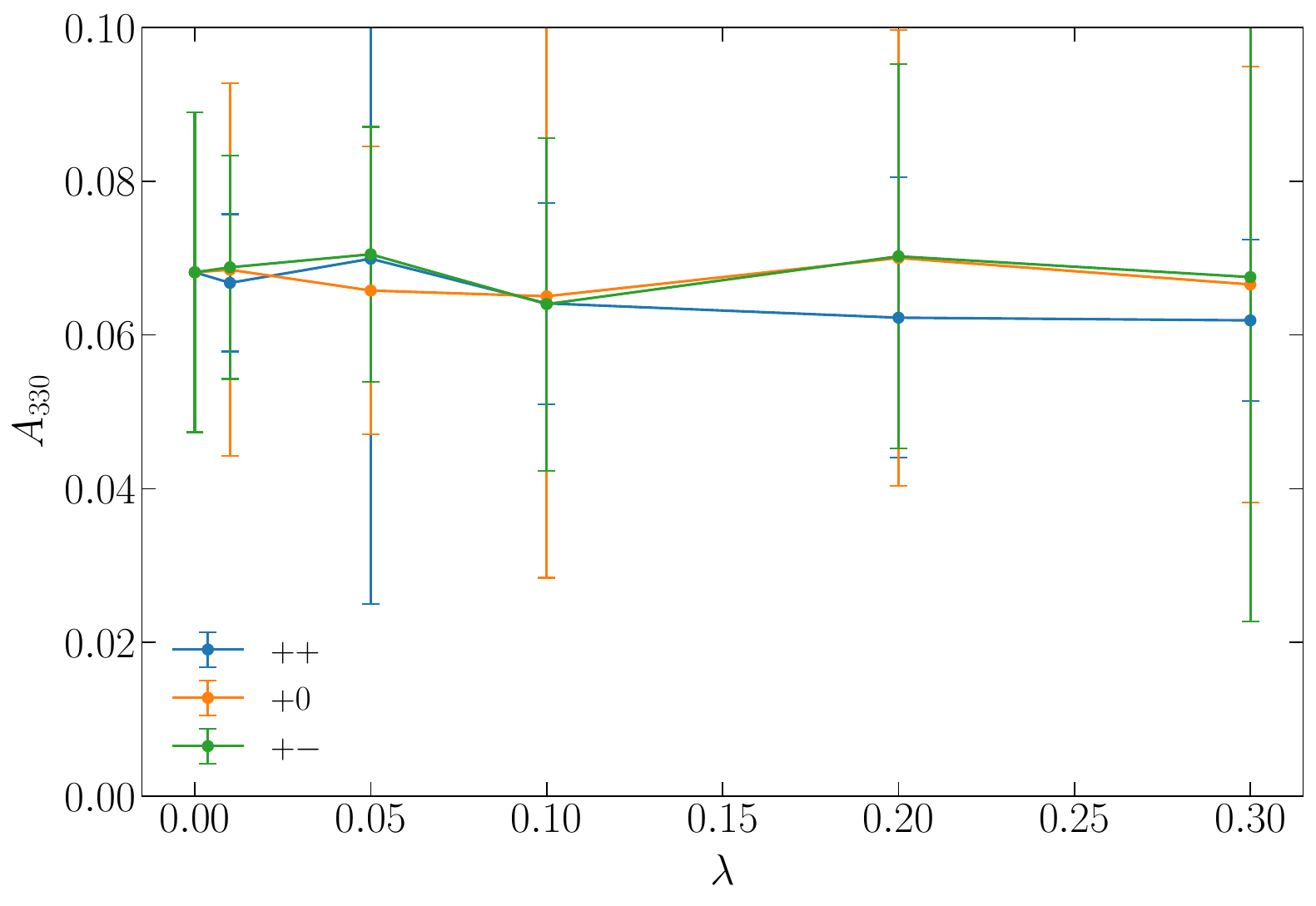}
  \caption{\label{fig:330_A} Similar to Fig.~\ref{fig:220_A}, but for $330$ mode. As the mass ratio of this system is only around $q\sim1.2$, the $330$ mode has a relatively small excitation amplitude.
  }
\end{figure}

In Figs.~\ref{fig:220_A},~\ref{fig:221_A} and~\ref{fig:330_A}, we show the amplitude extraction results for $220$, $221$, and $330$ modes respectively. Note that we have transformed the direct fitting results $B_k$ and $\varphi_k$ to $A_k$ and $\phi_k$ for comparison with previous studies. We estimate the uncertainty of the extracted amplitude via $\sigma_{A}\sim \sigma_f/|\omega|^2$, where $\omega=\omega_{r}+i\omega_i$ is the complex mode frequency. We shall note that $\sigma_A$ is only a rough estimator of the statistical uncertainty of the fitting result, and it may not reflect the true uncertainty for the mode extraction. It is clear that for the $221$ mode amplitude shown in Fig.~\ref{fig:221_A}, the fluctuation suggests a much larger uncertainty compared to that shown by the error bar. This is consistent with expectation as to extract the $221$ mode, we use a much shorter time window according to Eq.~(\ref{eq:time_window}), which leads to a small statistical uncertainty that underestimates the real uncertainty. A proper scaling with the length of the time window might provide a better estimation of the uncertainty. However, we should emphasize that the estimated statistical uncertainty is in principle different from the systematic uncertainty of the extraction procedure~\cite{Cheung:2023vki,Zhu:2024rej}.

From Fig.~\ref{fig:220_A} we can see that, for the  $++$ branch, there is a weak trend that, as the charge-to-mass ratio $\lambda$ increases, the amplitude of the $220$ mode decreases. However, the change is only around $5\%$ even for $\lambda$ as large as $0.3$. This trend is consistent with the result shown in Fig.~8 of Ref.~\cite{Bozzola:2021elc} that the total energy radiated away in pre-merger GW decreases by about $5\%$ to $10\%$ for this system configuration. Also, similar to the results in Ref.~\cite{Bozzola:2021elc}, for $+0$ and $+-$ branches, the amplitudes of the $220$ mode show no clear trend with the current extraction precision.

Combining the above three figures, we find that for the modes that dominate the ringdown emission, namely the $220$, $221$, and $330$ modes, their mode excitation amplitudes are weakly dependent on the charge-to-mass ratio of the progenitor BHs up to $\lambda=0.3$. The excitations of these QNMs are basically consistent with the expectations from Kerr BBH merger simulations~\cite{Cheung:2023vki} in agreement with the discussion in Ref.~\cite{Bozzola:2021elc}. Note that for the same simulations, there can be a significant accumulated phase difference during the inspiral, especially for large $|\lambda|\leq 0.3$~\cite{Bozzola:2021elc}. 
 
% the following discussion seems to be too detailed, let's wait for someone's asking
% We also compared our results with the fitting results given in Ref.~\cite{Cheung:2023vki}, which is for uncharged BBH mergers in the SXS catalog~\cite{Boyle:2019kee}. We find that our results presented here is generally consistent with their results considering the large uncertainties, but seems to be systematically a little larger than the fitting there. For $220$, $221$ and $330$ modes, the fitting from Ref.~\cite{Cheung:2023vki} gives the amplitude of about $0.97$, $4.1$ and $0.054$ respectively (we also notice that, in the jaxqualin website, for simulation SXS:BBH:1145, which is a event with symmetric mass ratio $\eta=0.2469$ that close to our simulations with $\eta=0.2471$, the mode amplitude is $0.97$, $4.4$ and $0.067$ respectively), smaller than our fitting for the GR case $\lambda=0$. This systematic larger might come from that we use the finite extraction radius, an interpolation to future null infinity seems to give a smaller overall amplitude. 

\begin{figure}[h]
  \centering
  \includegraphics[width=0.45\textwidth]{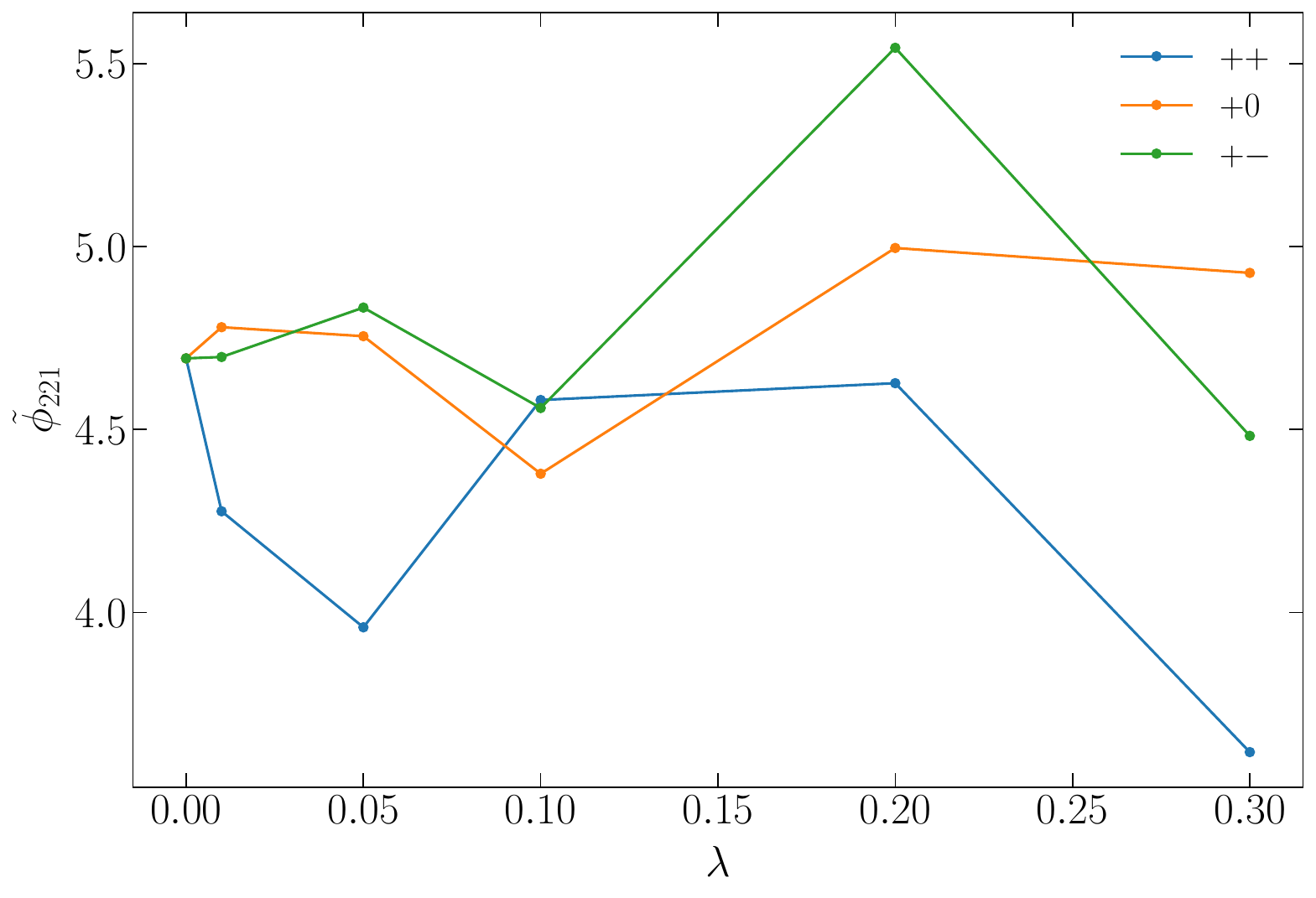}
  \caption{\label{fig:221_phi} The adjusted phase of $221$ mode, $\tilde{\phi}_{221}=2\phi_{221}-2\phi_{220}$. The phase of each mode is properly shifted by $2\pi$ so that the difference lies in a small interval. 
  }
\end{figure}

\begin{figure}[h]
  \centering
  \includegraphics[width=0.45\textwidth]{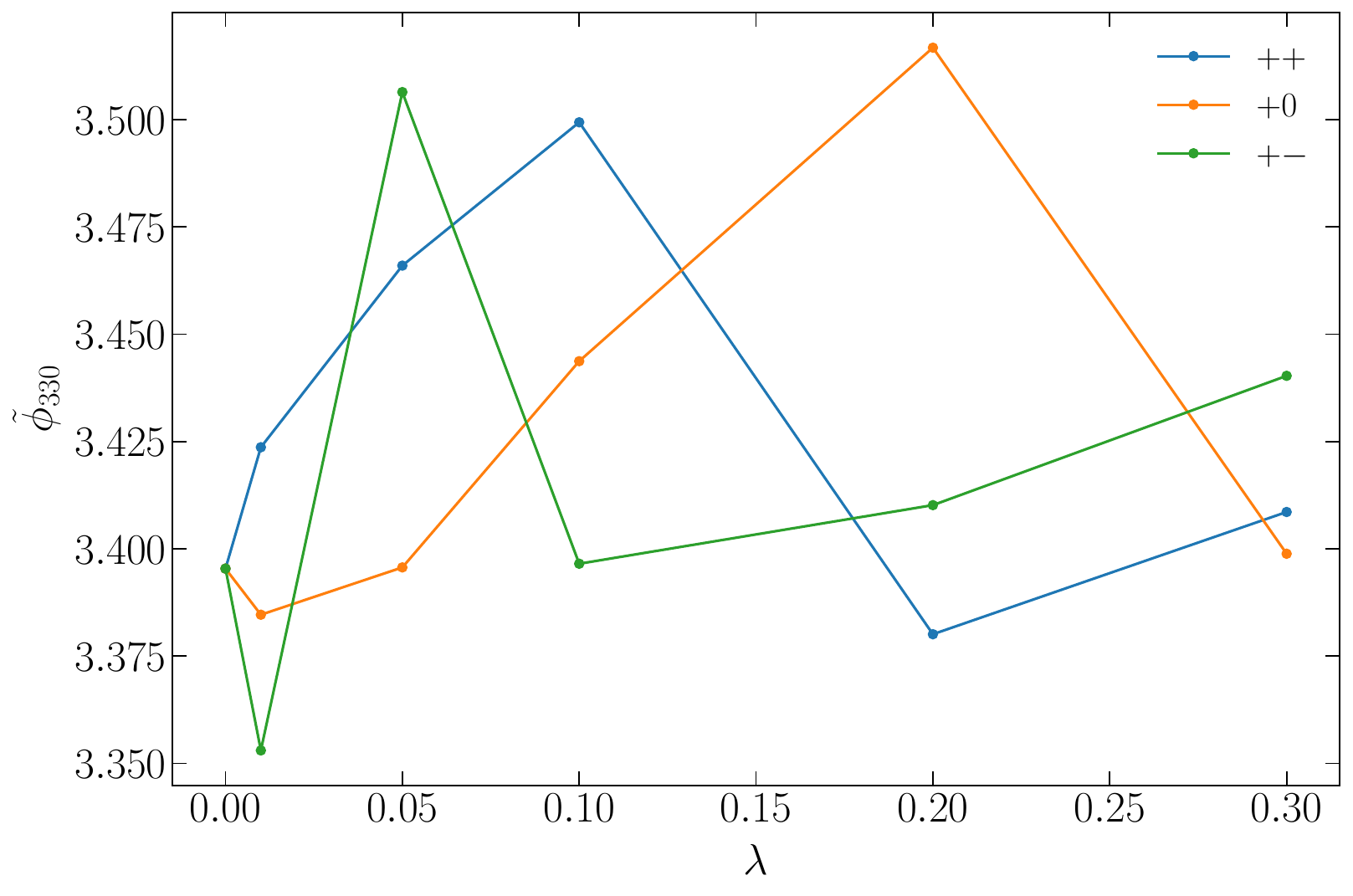}
  \caption{\label{fig:330_phi} The adjusted phase of $330$ mode, $\tilde{\phi}_{330}=2\phi_{330}-3\phi_{220}$. 
  }
\end{figure}

We also extract the mode phases at the reference time as shown in Fig.~\ref{fig:221_phi} and Fig.~\ref{fig:330_phi}. As the absolute phase of the mode is simulation dependent, we show the adjusted phase of the $221$ and $330$ modes defined by~\cite{Cheung:2023vki}
\begin{equation}
    \tilde{\phi}_{lmn}=2\phi_{lmn}-m\phi_{220}\,.
\end{equation}
Similarly to the mode amplitude, the adjusted phase of the modes shows no clear trend considering the extraction precision. Thus, the ringdown signal does not depend strongly on the initial charge of the BHs up to $\lambda=0.3$, for which the inspiral is already largely influenced. Considering our extraction precision, we may conclude that, for $\lambda$ as large as $0.3$, the mode amplitude changes less than $10\%$. For building a ringdown waveform, the mode amplitudes and phases from Kerr BBH merger are then a good approximation even if the final BH is charged and has a KN QNM spectrum, at least for $|\lambda| \leq 0.3$. In Table~\ref{tab:devi}, we compare the change in different quantities of the waveform for a charged BBH merger with $\lambda=0.3$ and the two initial BHs carry the same charge sign.

% \begin{table}[h]
%     \centering
%     \caption{The relative deviation from zero charge case for several different waveform quantities of the charged BBH merger with $\lambda=0.3$ in the $++$ situation. The last two lines also give the difference of the total energy emitted by GWs and the phase difference in the inspiral stage~\cite{Bozzola:2021elc}.}\label{tab:devi}
%     \renewcommand{\arraystretch}{1.2}
%     \begin{tabularx}{0.8\columnwidth}{>{\centering\arraybackslash}X>{\centering\arraybackslash}X}
%         \hline
%          $\left|({\rm KN}-{\rm Kerr})/{\rm Kerr}\right|$ & Relative deviation \\
%          \hline
%         $\delta\omega_r^{220}/\omega^{220}_{r,{\rm Kerr}}$ & $0.03$ \\
%         $\delta\omega_i^{220}/\omega^{220}_{i,{\rm Kerr}}$ & $0.01$ \\
%         $\delta A_{220}/A_{220,{\rm Kerr}}$ & $0.06$ \\
%         $\delta E^{\rm GW}_{f_0\rightarrow f_{\rm peak}}/E^{\rm GW}_{f_0\rightarrow f_{\rm peak},{\rm Kerr}}$ & $0.11$\\
%         $\delta\phi_{f_0\rightarrow f_{\rm peak}}/2\pi$ & $0.60$ \\
%         \hline
%     \end{tabularx}
% \end{table}

\begin{table}[h]
    \centering
    \caption{The relative deviation from zero charge case for several different waveform quantities of the charged BBH merger with $\lambda=0.3$ in the $++$ situation. The last two lines also give the difference of the total energy emitted by GWs and the phase difference in the inspiral stage~\cite{Bozzola:2021elc}.}\label{tab:devi}
    \renewcommand{\arraystretch}{1.2}
    \begin{tabular}{c c}
        \hline
         $\left|({\rm KN}-{\rm Kerr})/{\rm Kerr}\right|$ & Relative deviation \\
         \hline
        $\delta\omega_r^{220}/\omega^{220}_{r,{\rm Kerr}}$ & $0.03$ \\
        $\delta\omega_i^{220}/\omega^{220}_{i,{\rm Kerr}}$ & $0.01$ \\
        $\delta A_{220}/A_{220,{\rm Kerr}}$ & $0.06$ \\
        $\delta E^{\rm GW}_{f_0\rightarrow f_{\rm peak}}/E^{\rm GW}_{f_0\rightarrow f_{\rm peak},{\rm Kerr}}$ & $0.11$\\
        $\delta\phi_{f_0\rightarrow f_{\rm peak}}/2\pi$ & $0.60$ \\
        \hline
    \end{tabular}
\end{table}

\begin{figure}[h]
  \centering
  \includegraphics[width=0.45\textwidth]{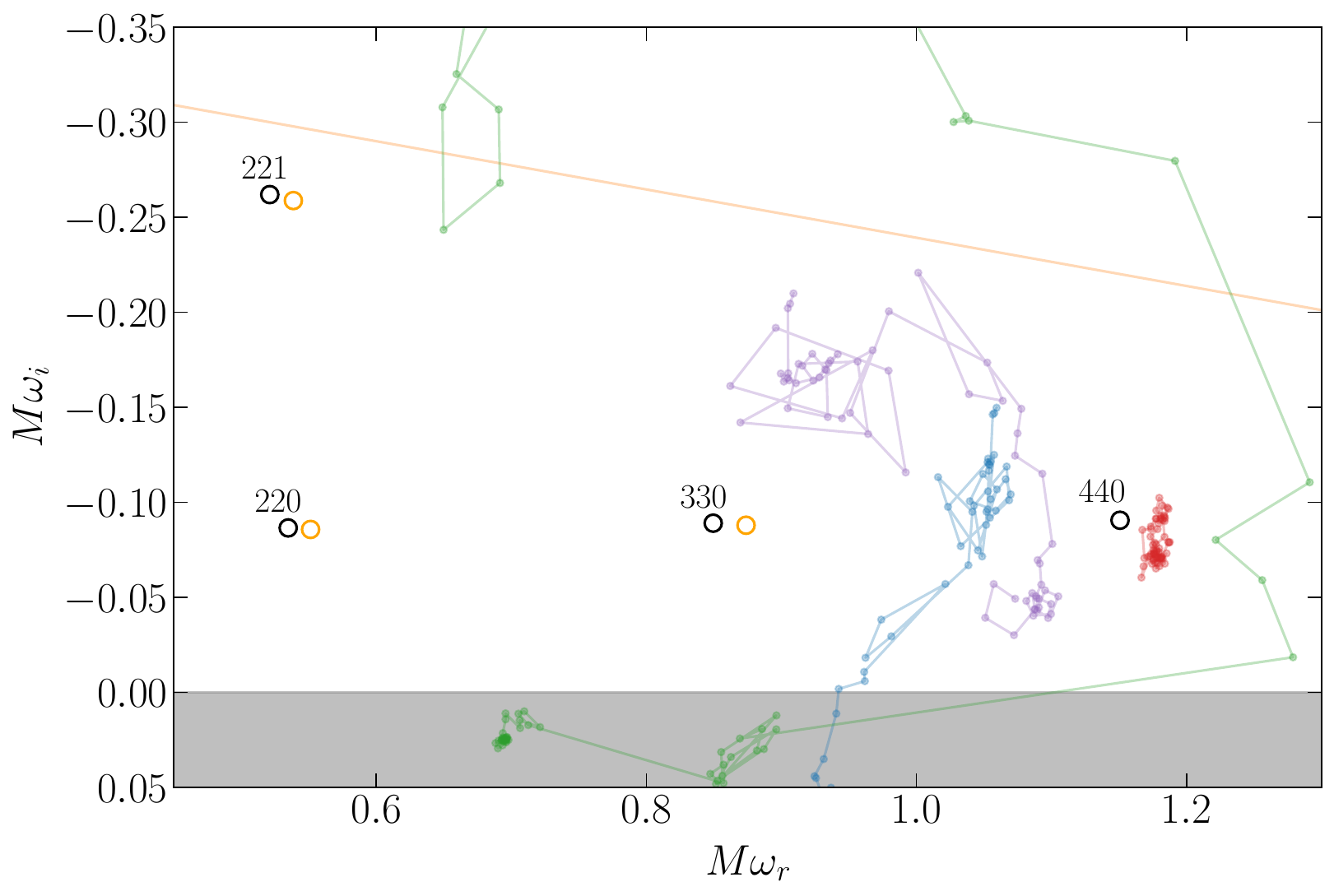}
  \caption{\label{fig:440_omega} The fitting results for the $440$ mode of $++$ simulation and $\lambda=0.3$ in the mode frequency plane. The black circles denote the mode frequencies of Kerr BH, and the orange circles denote the KN counterparts. The dots with the same color show the evolution of the fitting result of a single mode. The gray region corresponds to $\omega_i>0$, where the mode is exponentially growing.
  }
\end{figure}

We further perform mode extraction for those modes without a priori known frequencies from perturbative calculations. As discussed before, for these modes, we can only perform a fully frequency-agnostic fitting where all mode amplitudes, phases, and frequencies are treated as free parameters. The fitting results suggest the existence of some higher modes. For example, in Fig.~\ref{fig:440_omega} we show the fitting results of the $440$ mode of $++$ simulation and $\lambda=0.3$ in the mode frequency plane~\cite{Cheung:2023vki}. The dots connected by lines denote the fitted mode frequencies for different starting time $t_0$, while the black circles show the QNM frequencies of Kerr BH and orange circles show the corresponding KN counterparts, for those modes that have already been calculated \cite{Dias:2021yju,Dias:2022oqm}. It is clear that the red dots in the figure represent the existence of the $440$ mode. However, these higher modes are, in general, too weak (partially because of the mass ratio that is close to $1$) and due to the limited waveform quality, we are unable to provide a stable enough fitting for their excitation amplitudes and phases. 

%---------------------------------------------------------------------
\section{Future measurement prospect}\label{sec:det}
%---------------------------------------------------------------------

Having the numerically simulated waveforms of charged BBH mergers, we also investigate how well future GW observations can constrain the charge of the final BH from a ringdown-only analysis based on Bayesian inference. Previous studies~\cite{Carullo:2021oxn,Gu:2023eaa} generally use analytic waveform injections that consist only of two or three modes and ignore the nonlinear nature close to the merger time and the presence of other modes, although these effects might become relevant for high SNR events. We explore these aspects with a more realistic injection in this section.  

Our analysis of the KN ringdown is based on the {\tt pyRing} software, which is a time-domain Bayesian inference python package employed by the LIGO-Virgo-KAGRA (LVK) collaboration to perform ringdown-only tests of GR~\cite{Carullo:2019flw,carullo:2023}. The details of the time-domain analysis of the ringdown signal can be found, for example, in Ref.~\cite{Isi:2021iql}. Here we give only a brief overview. 

The log-likelihood function for the observed GW data $d(t)$, given the presence of a ringdown signal $h(t;\theta)$, is~\cite{Carullo:2019flw}
\begin{eqnarray}
    \ln p(d|\theta)&=&-\frac{1}{2}\int\int {\rm d}t {\rm d}t' (d(t)-h(t;\theta))C^{-1}(t-t')\nonumber\\
    &&\times (d(t')-h(t';\theta))\,,
\end{eqnarray}
where $C(\tau)$ is the autocovariance function of the detector noise $n(t)$, and $\theta$ are the parameters determining the signal $h$. Bayes' theorem states that the posterior distribution is given by 
\begin{equation}
    p(\theta|d)=\frac{p(\theta)p(d|\theta)}{p(d)}\,,
\end{equation}
where $p(\theta)$ is the prior of the parameters and $p(d)=\int p(\theta)p(d|\theta){\rm d}\theta$ is the so-called evidence. 

We take the numerically simulated waveform as zero-noise injection $d(t)$, and restrict our template for ringdown fitting to 
\begin{equation}
    h^+-ih^\times=\sum_{\substack{lmn=220,\\221,330}}\mathcal{A}_{lmn}e^{-i\omega_{lmn}(t-t_0)-i\Phi_{lmn}}\prescript{}{-2}{S}_{lmn}(\iota,\varphi)+\mathcal{M}.\,,
\end{equation}
where $\prescript{}{-2}{S}_{lmn}$ designates the spheroidal harmonics~\cite{Berti:2005gp}, $\iota$ and $\varphi$ are the spin polar angle and spin azimuthal angle, respectively. We always set $\varphi=0$ as it is completely degenerate with the overall mode phase. $t_0$ is the starting time of our fitting and $\mathcal{A}_{lmn}$ and $\Phi_{lmn}$ are the mode amplitude and phase reference to the starting time. $\mathcal{M}.$ represents the mirror modes and they are determined by the reflection (nonprecessing) symmetry~\cite{Berti:2007fi}. We only take into account the $220$, $221$, and $330$ modes in the template. The reason for this choice is
the availability of fitting formulae relating the mode frequency $\omega_{lmn}$ to the remnant BH mass $M_f$, spin $a_f$, and charge $Q_f$ in the literature for these modes~\cite{Carullo:2021oxn}. Further, we ignore the counterrotating modes as they are expected to be only significantly excited in systems with large anti-aligned progenitor spins to the orbital angular momentum~\cite{Cheung:2023vki}. 

To construct the observed GW data, the polarization angle $\psi$ and the orientation of the GW event relative to the detector must also be specified. We fix the sky location to the estimated value for the GW150914 event but keep $\psi$ as a free parameter. In summary, the parameters for our analysis include
\begin{equation}
    \{M_f,\chi_f,\lambda_f,\mathcal{A}_{lmn},\Phi_{lmn},\iota,\psi\}\,,
\end{equation}
where $\chi_f=J_f/M_f^2$ is the dimensionless spin and $\lambda_f=Q_f/M_f$ is the charge-to-mass ratio of the remnant BH. We use a uniform prior for $M_f$ in the range of $[10,200]\,M_\odot$, with $\chi_f$ and $\lambda_f$ uniformly distributed within the quarter disc defined by $\chi_f^2+\lambda_f^2<0.99$. $\cos\iota$ is chosen from $\mathcal{U}(0,1)$ and both $\Phi_{lmn}$ and $\psi$ have uniform prior $\mathcal{U}(0,2\pi)$. We set $\mathcal{A}_{220}\in[0,20]\times10^{-21}$, $\mathcal{A}_{221}\in[0,10]\times10^{-21}$, and $\mathcal{A}_{330}\in[0,0.8]\times10^{-21}$ uniformly according to our experiments. Although in the previous section we obtained the expected amplitudes of each mode, the range of the prior still needs to be adjusted to cover the posterior. The priors of these parameters are assumed to be independent of each other.

To construct $d(t)$ from $\Psi_4^{lm}$, we use the following procedure. We first obtain $h_{lm}$ by integrating $\Psi_4^{lm}$ with the fixed frequency integration method~\cite{Reisswig:2010di}. Based on the assumed initial mass and luminosity distance of the event, we rescale the waveform and then combine it with $\prescript{}{-2}{Y}_{lm}(\iota,\varphi)$ to obtain the waveform at the specific direction. Finally, using the python package \texttt{SWIGLAL}, we convert the waveform to the detector frame, which is taken as the input $d$ in the Bayesian analysis.

Since we use the simulation waveform as the injection, the reference time $t_0$ should not be simply set to the peak time of the strain, where the nonlinear effects are strong~\cite{Giesler:2019uxc}. Though in previous works for analyzing the charge detectability, $t_0$ is chosen to be $t_{\rm peak}$ even for real GW data~\cite{Carullo:2021oxn,Gu:2023eaa}, we find this leads to an unstable result of $\chi_f-\lambda_f$ estimation that strongly depends on the phase of the waveform at the reference time. Thus, in our following analysis, we fix $t_0=t_{\rm peak}+10\,M$. The proper choice of the starting time is still under investigation~\cite{Giesler:2019uxc}, and a late starting time could provide a clean ringdown signal but will further suppress the SNR. The choice used here might not be the most optimal one, but we leave this problem for future study. 

\begin{figure*}[t]
  \centering
  \includegraphics[width=0.85\textwidth]{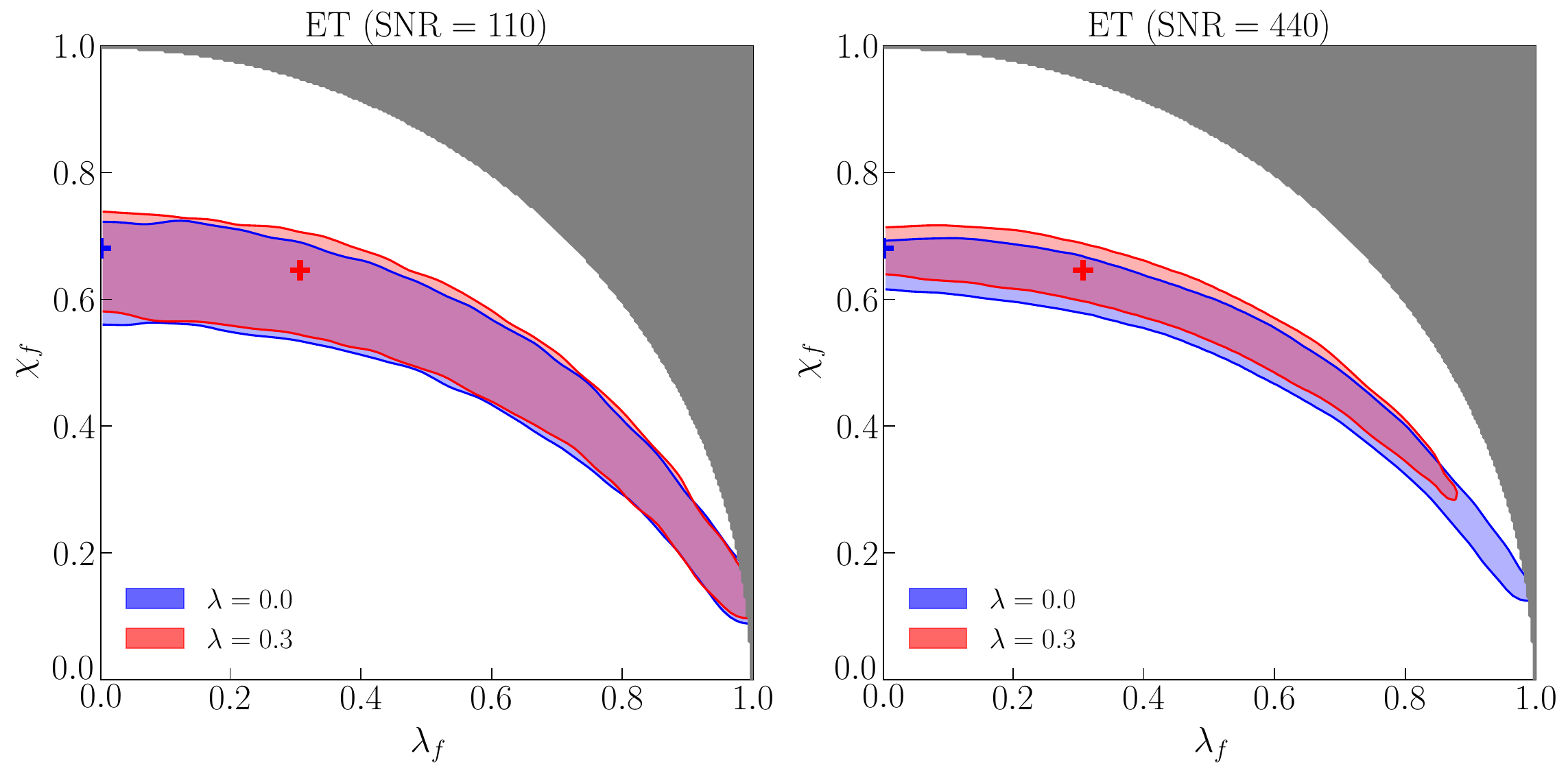}
  \caption{\label{fig:chi_q_SNR} Charge-spin posterior distribution for a GW150914-like event to be detected by ET for two different initial charge-to-mass ratios and two different SNRs. The true values of the remnant spin and charge are read out from the simulation data and marked with a ``$+$'' in the figure. For $\lambda=0$, $(\lambda_f,\chi_f)=(0,0.680)$ and for $\lambda=0.3$, $(\lambda_f,\chi_f)=(0.307,0.646)$. Solid filled contours are the $90\%$ credible region while the gray region violates the constraint $\chi_f^2+\lambda_f^2<0.99$. The left panel with lower SNR corresponds to an event that happens at the same location as GW150914, while in the right panel the SNR is larger as we assume a four times closer GW source.
  }
\end{figure*}

\begin{figure}[h]
  \centering
  \includegraphics[width=0.45\textwidth]{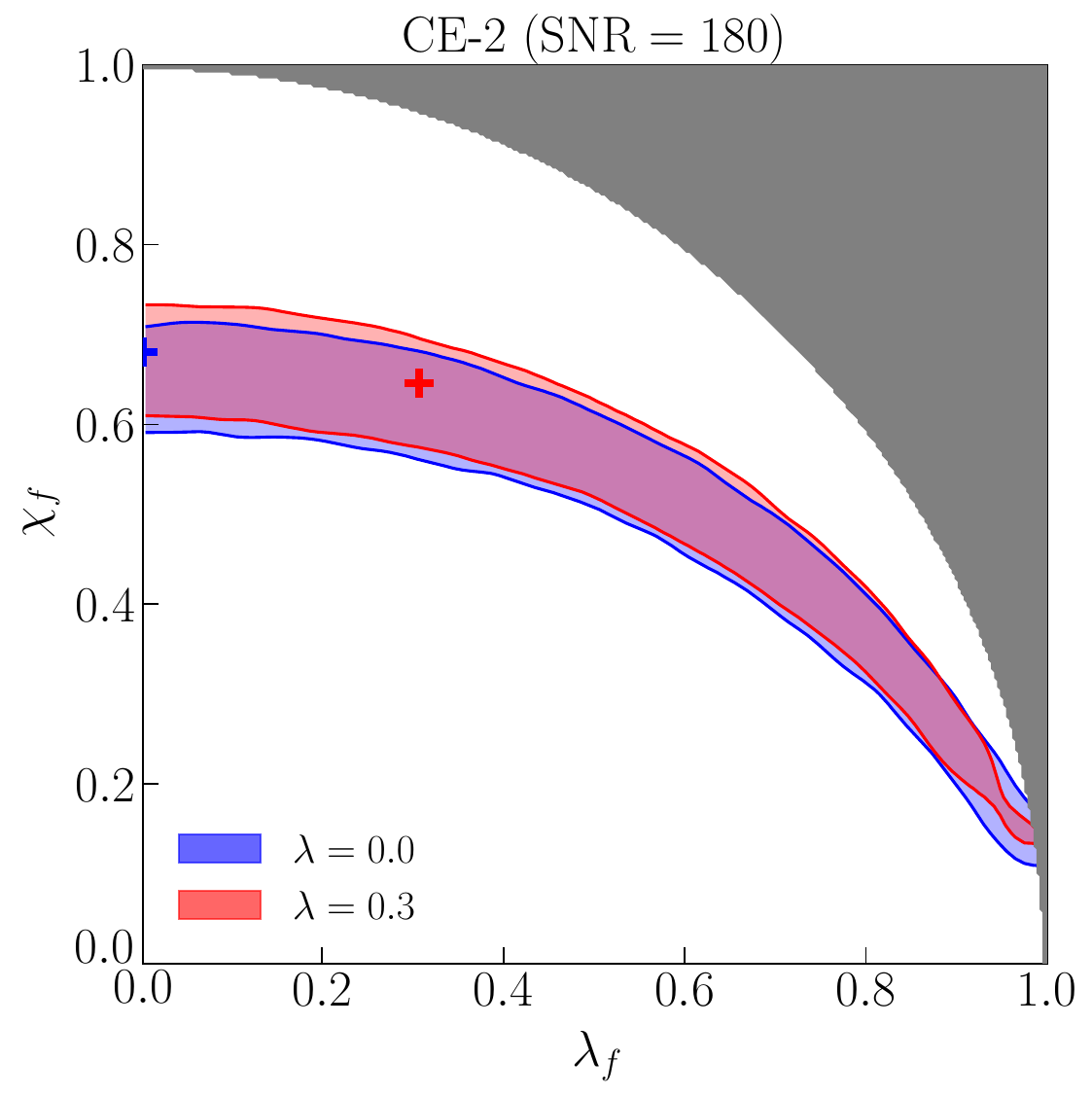}
  \caption{\label{fig:chi_q_330_CE} Similar figure as before but for CE. The ringdown SNR of a GW150914-like event observed by CE is slightly larger than that for ET, but there is no significant difference in the parameter estimation results.
  }
\end{figure}

Given that only limited information can be extracted from observations with current detectors~\cite{Carullo:2021oxn,Gu:2023eaa}, we consider the detectability of BH charge for a GW150914-like event using the noise curves of the next-generation GW detectors ET~\cite{Punturo:2010zz,Hild:2010id,Abac:2025saz} and CE~\cite{Reitze:2019iox,Reitze:2019dyk}. These planned next-generation GW observatories are designed to achieve a tenfold enhancement in sensitivity compared to LIGO detectors. We use the ET-D and CE-2 noise spectrum provided by \texttt{GW-Toolbox}~\cite{Yi:2021wqf}. In Fig.~\ref{fig:chi_q_SNR}, we show the charge-spin posterior distribution for the GW150914-like event detected by ET but for two different initial charge-to-mass ratios $\lambda=0$ and $\lambda=0.3$. We consider the $++$ simulation so that the remnant BH can have the largest charge. The SNR for our injected event is about $110$ if we take the total progenitor mass to be $70.5\,M_\odot$ and the luminosity distance $D_L=410\,{\rm Mpc}$. We also present a result that increases the SNR by a factor of four, corresponding to a GW150914-like event happening at a distance of $D_L\sim 100\,{\rm Mpc}$. The contours in the figure show the $90\%$ credible region and the true value of $\chi_f$ and $\lambda_f$ are marked by the ``$+$'' symbol in the figure. We note that, due to the different initial charge, the final dimensionless spin of the BH is different for the two simulations~\cite{Bozzola:2021elc}. 
Figure~\ref{fig:chi_q_330_CE} shows a corresponding case detected by CE. The SNR is slightly higher than that in the ET case, but there is no significant difference since ET and CE have similar noise curves. Considering these similarities, in the following discussion, we use ET as the representative case. 

%----------------

The results shown in Fig.~\ref{fig:chi_q_SNR} suggest that previous estimations of charge detectability in Ref.~\cite{Gu:2023eaa}, according to which one can obtain a constraint $Q<0.2$ at $90\%$ confidence level with ET for a GW150914-like event, might be too optimistic. Starting from a relatively late time is required in realistic ringdown analysis due to the non-linear spacetime response around the peak time. In fact, our choice of $t_0=10\,M$ should be considered as a conservative choice in the sense that, for only fitting the first overtone $221$, starting from $t_0\sim15-20\,M$ is sufficient~\cite{Giesler:2019uxc}. Considering that at least two modes are needed to break the charge-spin degeneracy, $220$ and $221$ modes have to be measured in this case (the $330$ mode seems to be too weak in this system, as discussed later). However, the $221$ mode decays much faster than the fundamental mode $220$, thus starting from late time effectively decreases the amplitude of the higher overtone, making it harder to constrain the charge. When we further increase the SNR, the contour for charged BHs shifts upward, which is consistent with the results shown in Ref.~\cite{Carullo:2021oxn}. We should also note that,  when numerical relativity simulations are employed it turns out that the spin of the remnant BH tends to be smaller for a larger charge-to-mass ratio in the $++$ case, which was not taken into account in previous analytical injection studies. 

\begin{figure}[h]
  \centering
  \includegraphics[width=0.45\textwidth]{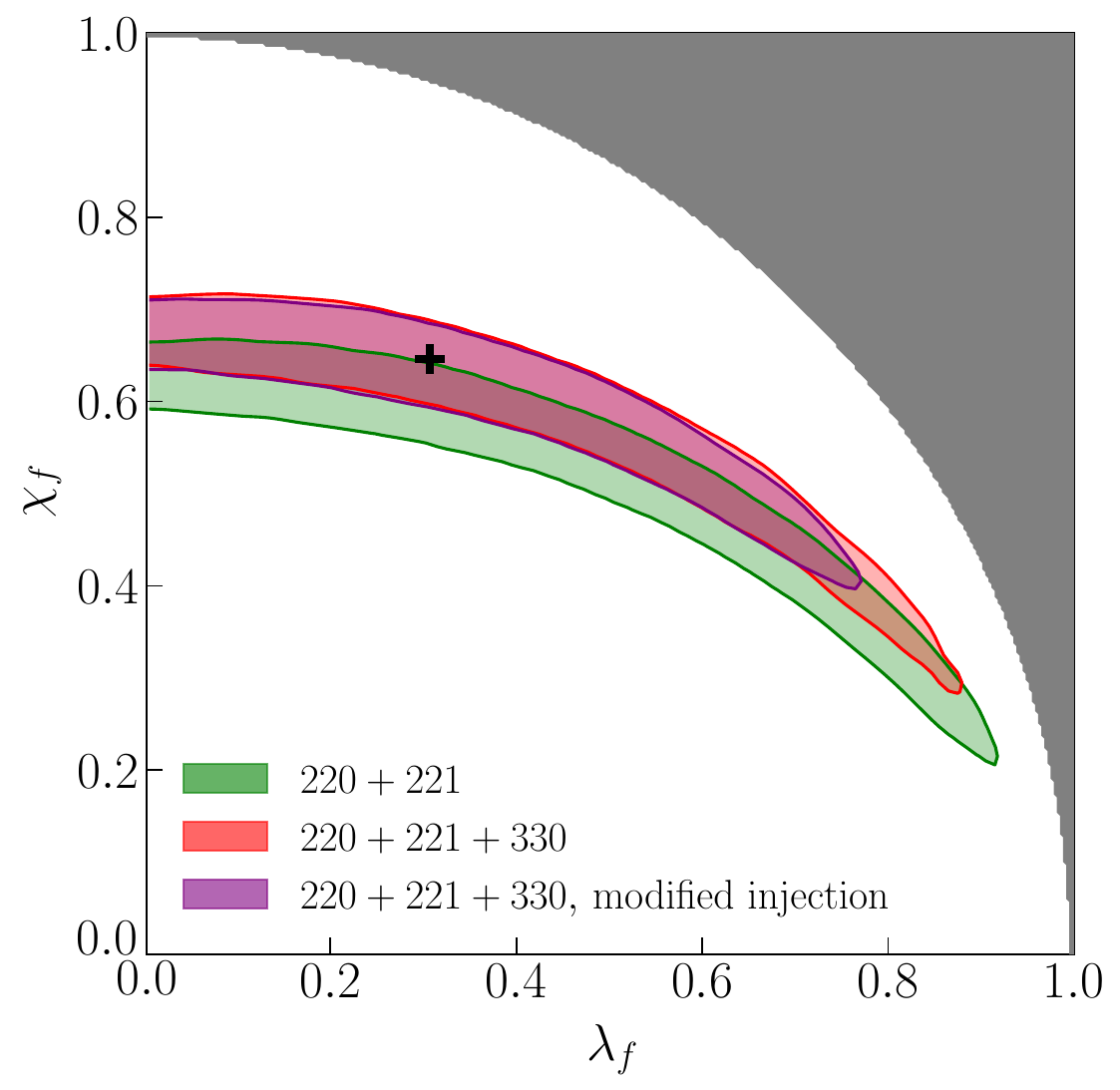}
  \caption{\label{fig:chi_q_330} Charge-spin posterior distribution for a GW150914-like event with $\lambda=0.3$ observed by ET. We use two waveform templates that both consider the $220$, $221$ modes but one also includes the $330$ mode. We further artificially increase the amplitude of the $330$ mode to show its effect on the parameter estimation. The employed SNR is roughly $440$. 
  }
\end{figure}

In Ref.~\cite{Gu:2023eaa}, it is suggested that the detection of the $330$ mode could help break the charge-spin degeneracy. However, the system considered in our simulation has a mass ratio around $1.25$, thus the excitation of the $330$ mode is relatively weak as shown in Sec.~\ref{sec:extraction result}. Nevertheless, in Fig.~\ref{fig:chi_q_330}, we show the comparison of the results using a template that only includes the $220$ and $221$ modes, and a template that further includes the $330$ mode. Since the considered ${\rm SNR}=440$ is rather large, ignoring the contribution from $330$ mode leads to a clear bias in the parameter estimation result as shown in this figure. However, it seems that including the $330$ mode decreases the bias but it does not contribute a lot to the charge-spin degeneracy breaking due to the small amplitude of this mode. In the figure, we also show a result where we artificially increase the amplitude of the $330$ mode by a factor of $2.5$ when we construct the numerical waveform. This roughly corresponds to a $330$ mode excitation in a system with a mass ratio around $2$~\cite{Cheung:2023vki}. The contour shows that the degeneracy between the charge and spin is lessened, consistent with the prediction in Ref.~\cite{Gu:2023eaa}. We conclude that, for a system with a larger $330$ mode amplitude, which happens when the progenitor's mass ratio is significantly different from unity, including the $330$ mode in the template can improve the constraint on the BH charge. However, even for a GW150914-like system with a mass ratio close to $1$ and for a large SNR as we considered here, including the $330$ mode is necessary for obtaining a less biased parameter estimation result.

\begin{figure}[t]
  \centering
  \includegraphics[width=0.45\textwidth]{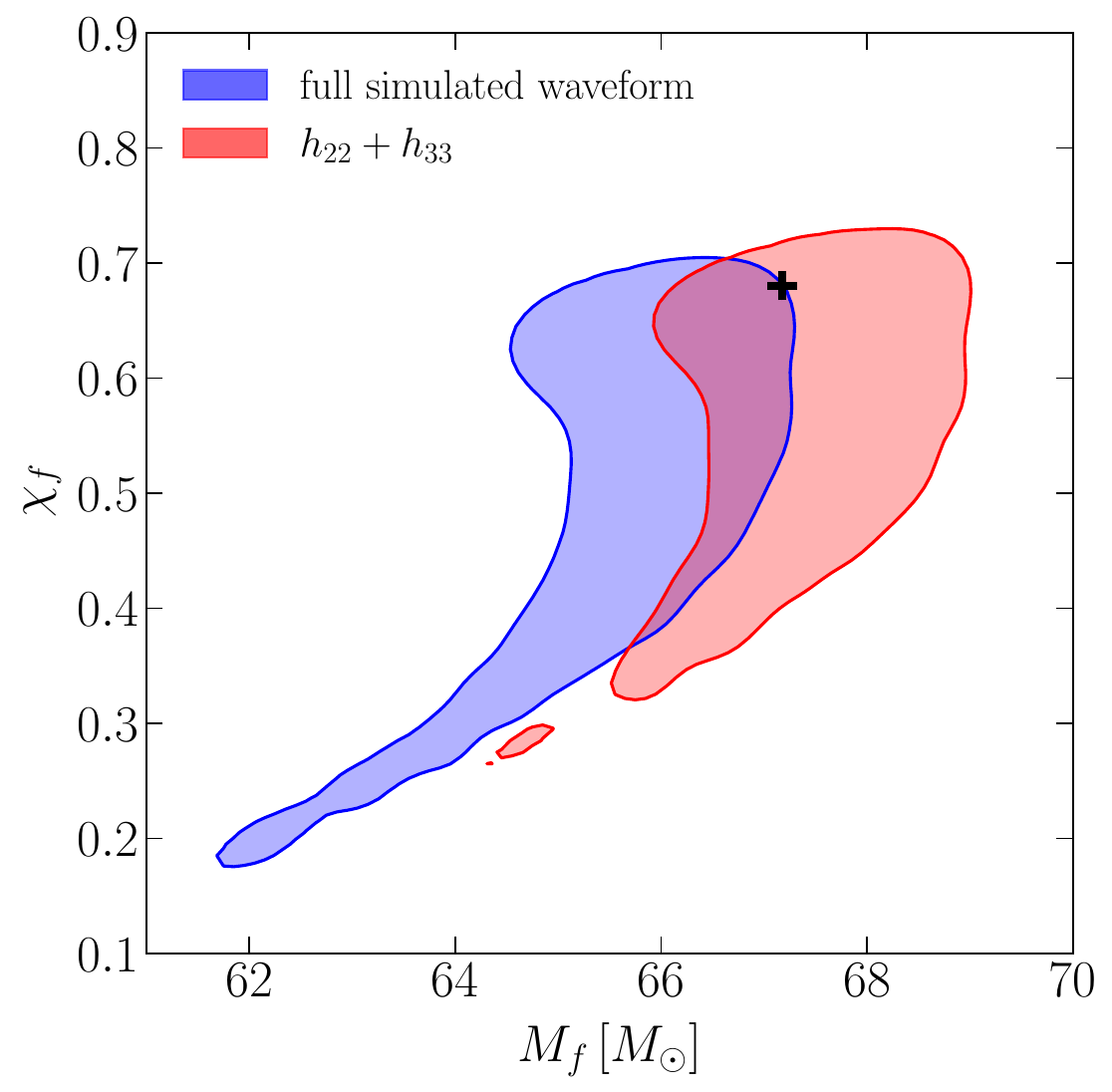}
  \caption{\label{fig:M_q} Mass-spin posterior distribution for a GW150914-like event with $\lambda=0$ observed by ET. To study the effects of higher modes, we construct two injections:  one is the full numerical simulated waveform and the other only takes into account the $l=|m|=2$ and $l=|m|=3$ components. The employed SNR is roughly 440.
  }
\end{figure}

We also notice that it might be necessary to consider even higher modes, like $440$ mode, in future data analysis, which requires the numerical calculation and interpolation of the spectrum of these modes through perturbative techniques, as in Refs.~\cite{Dias:2021yju,Dias:2022oqm}. Similarly to the $330$ mode, including these modes might not contribute a lot to the breaking of the charge-spin degeneracy considering their small amplitude, but ignoring them might lead to a biased result in parameter estimation. In Fig.~\ref{fig:chi_q_SNR}, one might notice that the spin estimation of the zero charge case seems to be biased from the true value even limited to the slice $\lambda_f=0$, and it becomes clearer with the increase of the SNR. 

Due to the lack of interpolation of the KN QNM frequencies for other modes, we investigate their influence via the following method. We construct a waveform from the numerical simulation data by only including the $l=|m|=2$ and $l=|m|=3$ components and perform a parameter estimation. In Fig.~\ref{fig:M_q}, we show the mass-spin posterior distribution of the full waveform injection and the injection that only includes $h_{22}$ and $h_{33}$ for the $\lambda=0$ case. It is clear that the parameter estimation for the full waveform when including the higher modes is biased from the true value, while for the artificially constructed waveform the parameter estimation is consistent. This indicates the necessity of including higher modes in the ringdown analysis even in the presence of a large charge-spin degeneracy that disturbs the parameter estimation. We note that this bias should also happen to the $\lambda=0.3$ case in Fig.~\ref{fig:chi_q_SNR} but it might be less noticeable in the figure due to the shape of the contour caused by the charge-spin degeneracy. 

%---------------------------------------------------------------------
\section{Conclusions}\label{sec:con}
%---------------------------------------------------------------------

In this study, we investigate the ringdown mode excitation of charged BBH mergers with initial charge-to-mass ratios up to $0.3$. We present the first fitting results of the amplitudes and the phases of the $220$, $221$, and $330$ modes. Using waveforms generated by simulations in full Einstein-Maxwell theory, we also forecast the charge measurability from ringdown-only observations for the future next-generation GW detectors ET and CE. 

The QNM extraction results suggest that for a GW150914-like event, the dominant ringdown mode excitation, namely, the excitation of the $220$, $221$, and $330$ modes depends only mildly on the charge-to-mass ratio of the progenitor BHs up to charge-to-mass ratio $\lambda=0.3$. The mode excitation is also not sensitive to how the progenitor BHs carry the charge, i.e., the excitation is similar for the $++$, $+0$, and $+-$ simulations. The maximum change of the $220$ mode amplitude in our results is less than $10\%$, and the trend is consistent with the analysis of the pre-merger GW signal~\cite{Bozzola:2021elc}, according to which for two initial BHs having the same sign of charge, the total energy emitted via GWs slightly decreases as the charge-to-mass ratio increases. The insensitivity of the mode excitation ratio to the initial charge-to-mass ratio probably suggests that the initial perturbation for the ringdown does not deviate much from the non-charged case in these quasi-circular simulations with charge. The main difference is only an overall phase shift. Further investigation of eccentric mergers in Einstein-Maxwell theory, or mergers with a higher charge-to-mass ratio, might be interesting. 

Based on the quality of the employed simulations \cite{Bozzola:2021elc} and the extraction procedure, we are unable to stably extract the higher mode amplitudes and phases. Future more accurate simulations and calculation of other mode frequencies of KN BHs through perturbative approaches, will enable further study. Nevertheless, our result already shows that using the QNM excitation predicted from a Kerr BBH merger is a reasonable starting point, at least for $|\lambda|<0.3$ initial charge-to-mass ratio. This can be important for building surrogate waveforms that consistently include the inspiral-merger-ringdown part of a charged BBH merger.

For studying the charge detectability of future GW detectors via ringdown-only analysis, we perform the first Bayesian inference that adopts an injection of a realistic waveform of a charged BBH merger. Our results indicate that previous studies~\cite{Carullo:2021oxn,Gu:2023eaa} might overestimate the charge detectability mainly due to the improper choice of the starting time of the ringdown analysis. Due to the non-linear nature around the peak time, it might be necessary to choose a late starting time for a clean ringdown signal and unbiased parameter estimation result. However, to break the charge-spin degeneracy, one needs to measure at least two modes independently. The late starting time will make the measurements of higher overtones harder and lead to a much worse constraint on the charge of the remnant BH. As pointed out in previous studies, we also verified that the measurement of the $330$ mode can help break the degeneracy, though for moderate mass ratio, e.g. up to 2, the improvement is only mild.  Another important observation from our results is that,  even if including these higher modes in the waveform cannot largely improve the charge measurement, they are still necessary in order to obtain an unbiased parameter estimation result for high SNR events. The mass-spin measurement can be significantly biased when ignoring these higher modes in the waveform template. Future studies in this direction will rely on the calculation and interpolation of higher mode frequencies of KN BHs.

Finally, as discussed in Ref.~\cite{Bozzola:2021elc}, apart from the fact that the remnant BH may carry charge, the mass and the spin of the remnant BH will also differ from a zero-charge BBH merger, especially in the $+-$ case, where the remnant BH would have a charge close to zero. In the sense of an inspiral-merger-ringdown consistency check, the mass and spin of the remnant BH can also provide valuable information that can be used to constrain the charge-to-mass ratio of the progenitor BHs, which is not studied in the current paper but remains an important topic for future investigations. 

%---------------------------------------------------------------------
\acknowledgments
%---------------------------------------------------------------------
We would like to thank Pierre Mourier, Adrian Kuntz, Lorena Zertuche, Marc Besancon, and Nicola Franchini for useful discussions.
Z.H., Z.W., and L.S.\ are supported by the National Natural Science Foundation of China (124B2056, 123B2043), the Beijing Natural Science Foundation (1242018), the National SKA Program of China (2020SKA0120300), and the Max Planck Partner Group Program funded by the Max Planck Society. Z.H. is supported by the China Scholarship Council (CSC).
This work was in part supported by NASA grant 80NSSC24K0771 and NSF grant PHY-2145421 to the University of Arizona. This research is part of the Frontera computing project at the Texas Advanced Computing Center. Frontera is made possible by U.S. National Science Foundation award OAC-1818253. This work used Stamepede3 at the Texas Advanced Computing Center through allocation PHY190020 from the Advanced Cyberinfrastructure Coordination Ecosystem: Services \& Support (ACCESS) program, which is supported by U.S. National Science Foundation grants 2138259, 2138286, 2138307, 2137603, and 2138296~\citep{boerner_access_2023}.  
The partial support of KP-06-N62/6 from the Bulgarian science fund is also gratefully acknowledged. DD acknowledges financial support via an Emmy Noether Research Group funded by the German Research Foundation (DFG) under grant no. DO 1771/1-1, and the Spanish Ministry of Science and Innovation via the Ram\'on y Cajal programme (grant RYC2023-042559-I), funded by MCIN/AEI/ 10.13039/501100011033. We acknowledge Discoverer PetaSC and EuroHPC JU for awarding this project access to Discoverer supercomputer resources.

%---------------------------------------------------------------------
\bibliography{refs.bib}
%---------------------------------------------------------------------

\end{document}